\documentstyle[12pt,fleqn]{article}

\catcode`\@=11
\@addtoreset{equation}{section}

\begin{document}
\begin{flushright}
OU-HET/201\\hep-th/9409???\\September 1994
\end{flushright}
\vspace{0.5in}
\begin{center}\Large{\bf Chern-Simons Quantization of
(2+1)-Anti-De Sitter Gravity on a Torus} \\
\vspace{1cm}\renewcommand{\thefootnote}{\fnsymbol{footnote}}
\ Kiyoshi Ezawa\footnote[1]{Supported by JSPS.}
\footnote[2]{e-mail address:
ezawa@funpth.phys.sci.osaka-u.ac.jp}\setcounter{footnote}{0}

\vspace{0.5in}

        Department of Physics \\
        Osaka University, Toyonaka, Osaka 560, Japan\\
\vspace{0.1in}
\end{center}
\vspace{0.5in}
\baselineskip 17pt
\begin{abstract}
Chern-Simons formulation of 2+1 dimensional Einstein gravity
with a negative cosmological constant is investigated when the
spacetime has the topology ${\bf R}\times T^{2}$. The physical
phase space is shown to be a direct product of two sub-phase
spaces each of which is
a non-Hausdorff manifold plus a set with nonzero
codimensions.
Spacetime geometrical interpretation of each point in the
phase space is also given and we explain the 1 to 2
correspondence with the ADM formalism from the geometrical
viewpoint. In quantizing this theory, we construct a
\lq\lq modified phase space" which is a cotangnt bundle
on a torus. We also provide a modular invariant inner product
and investigate the relation to the quantum theory which is
directly related to the spinor representation of
the ADM formalism.
(This paper is the revised version of a previous
paper(hep-th/9312151). The wrong discussion on the topology
of the phase space is corrected.)\\
                    \\
classification numbers (PACS nos.): 04.20, 04.60
\end{abstract}
\newpage
%%%%%%%%%%%%%%%%%%%%%
%%%%%%%%%%%%%%%%%%%%%

\baselineskip 20pt

\section{Introduction}

 \ \ \ \ Since the first order formalism  of 2+1 dimensional
Einstein gravity was shown to be equivalent to the Chern-Simons
gauge theories with noncompact gauge groups\cite{witte}\cite{town},
many works have appeared on this \lq\lq Chern-Simons gravity"(CSG).
Particularly in the case where the spacetime topology
is ${\bf R}\times T^{2}$ and the cosmological constant vanishes,
various aspects of CSG including its geometrical interpretation
and the structure of its phase space seem to have been elucidated
\cite{carli} \cite{ander} \cite{carl2} \cite{louko} \cite{unruh}.

As for the case with nonvanishing cosmological constant, except
a series of works on the holonomy algebra which are made by
Nelson and Regge \cite{nelson}\cite{NRZ}\cite{CN}, relatively few
people deal with this case\cite{cenalo}.
In a previous paper\cite{ezawa}, we have
shown that the physical phase space in the negative cosmological
constant case has nine sectors when the spacetime has the topology
${\bf R}\times T^{2}$ and one of these sectors is in 1 to 2
correspondence with the ADM phase space. However we knew
little about the remaining eight sectors.

In this paper, we will give the topological and symplectic
structures to the whole of this phase space.
We find that this phase space is not equipped with
a cotangent bundle structure, and that the topological structure
is ditcinct according to
whether we take the gauge group to be ${\rm SO}(2,1)_{0}
\times{\rm SO}(2,1)_{0}$ or $\widetilde{\rm SL}(2,{\bf R})\times
\widetilde{\rm SL}(2,{\bf R})$.
We will give geometrical interpretations to each of the nine
sectors in the  ${\rm SO}(2,1)_{0}\times{\rm SO}(2,1)_{0}$ case.
Since the phase space does not have a cotangent bundle structure or
a real polarization, we cannot naively apply the conventional
quantization pocedure in which quantum states are represented by
functions of \lq\lq coordinates". To quantize such a phase space,
we need in general the help of geometric quantization \cite{wood}.
The discussion based on this geometric quantization, however,
tends to be abstruct. To make more concrete discussions be possible,
we modify the phase space
so that it can be a cotangent bundle. On the resulting phase space
we can use the conventional procedure of canonical quantization.

In \S 2  we briefly review the Chern-Simoms formulation of anti-de
Sitter gravity in the general case.
We also explain how to reduce the phase space and
how to obtain the symplectic structure of the reduced phase  space.
In \S 3, we investigate the phase space in the case where the
spacetime
is homeomorphic to ${\bf R}\times T^{2}$. We give a new
parametrization
in terms of which the nine sectors which have already appeared in
\cite{ezawa} can be described together.
The relation of the new parameter with the other observables which
were used in the previous works\cite{nelson}\cite{NRZ}\cite{CN}
\cite{carli}\cite{ezawa} are shown in \S 4. \S 5 is devoted to the
interpretation of the whole phase space in terms of the spacetime
geometry. The 1 to 2 correspondence is also explained in the
viewpoint
of the spacetime geometry. In \S 6, we give a quantization
prescription
using the new parameres as basic variables. Transformation properties
of
the new variables under the modular group are also investigated both
in
the calssical and quantum theories. \S 7 is devoted to the discussion
on the remaining issues.

Here we give the convention for the indices and the signatures of the
metrics used in this paper:
\begin{enumerate}
\item  $\mu,\nu,\rho,\cdots$ denote 2+1 dimensional spacetime indices
and  the metric $g_{\mu\nu}$ has the signature $(-,+,+)$.
\item $i,j,k,\cdots$ are  used for spatial indices.
\item $a,b,c,\cdots$ represent indices of the
SO$(2,1)$ vector representation
of the local Lorentz group, with the metric $\eta_{ab}=
{\rm diag}(-,+,+)$.
\item $\hat{a},\hat{b},\hat{c},\cdots$ denote indices of the
SO$(2,2)$ vector representation of the anti-de Sitter group.
The metric is given by $\eta_{\hat{a}\hat{b}}=
{\rm diag}(-,+,+,+)$.
\item $\epsilon_{abc}$ is the totally antisymmetric pseudo-tensor
with
$\epsilon_{012}=-\epsilon^{012}=1$.
\item $\epsilon^{ij}$ denotes the totally antisymmetric tensor
density on the spatial hypersurface $\Sigma$ with $\epsilon^{12}=1$.
\end{enumerate}

%%%%%%%%%%%%%%%%%%%%%%%%%%%%%%%%%%%%%%%
\section{Reduced Phase Space of Chern-Simons Gravity}

We start with the first-order gravity in (2+1)-dimensions with a
negative
cosmological constant $\Lambda=-1/L^{2}$. We use as fundamental
variables
the triad $e^{a}=e_{\mu}^{a}dx^{\mu}$ and the spin connection
$\omega^{ab}=\omega_{\mu}^{ab}dx^{\mu}$. The action is written as
\begin{eqnarray}
I_{EP} &=& \int_{M}\epsilon_{abc}e^{a}\wedge
[d\omega^{bc}+\omega^{b}_{\mbox{ }d}\wedge\omega^{dc}-
\frac{1}{3}\Lambda e^{b}\wedge e^{c}]		\nonumber \\*
&=&L\int_{M}E^{a}(2d\omega_{a}
+\epsilon_{abc}\omega^{b}\omega^{c}
+\frac{1}{3}\epsilon_{abc}E^{b}E^{c}),   \label{eq:EPac}
\end{eqnarray}
where we have introduced new variables $E^{a}\equiv\frac{1}{L}e^{a}$
and
$\omega^{a}\equiv\frac{1}{2}\epsilon^{a}_{\mbox{ }bc}\omega^{bc}$.

If we introduce the following (anti-)self-dual SO(2,1) connection
$$A^{(\pm)a}\equiv\omega^{a}\pm E^{a}, $$
the action (\ref{eq:EPac}) becomes the sum of two SO(2,1)
Chern-Simons actions
\begin{eqnarray}
I_{EP}&=&\frac{L}{2}\int_{M}(\eta_{ab}A^{(+)a}
\wedge dA^{(+)b}+\frac{1}{3}
\epsilon_{abc}A^{(+)a}\wedge A^{(+)b}\wedge A^{(+)c}) \nonumber \\*
& &-\frac{L}{2}\int_{M}(\eta_{ab}A^{(-)a}\wedge dA^{(-)b}+\frac{1}{3}
\epsilon_{abc}A^{(-)a}\wedge A^{(-)b}\wedge A^{(-)c})
\label{eq:CSWac}
\end{eqnarray}
up to surface terms.
To proceed to the canonical formalism a la Witten,
we assume that the spacetime
$M$ to be homeomorphic to ${\bf R}\times\Sigma$, where $\Sigma$ is a
two dimensional manifold \footnote{To simplify the analysis, we
assume
that $\Sigma$ is compact and has no boundary.},
and we naively set $x^{0}=t$
\begin{eqnarray}
I_{W}&=&(I_{EP})_{|M\approx R\times\Sigma} \nonumber \\*
&=&\int dt\int_{\Sigma}d^{2}x(-\frac{L}{2}\epsilon^{ij}\eta_{ab}
A_{i}^{(+)a}\dot{A}_{j}^{(+)b}+A_{t}^{(+)a}
{\cal G}_{a}^{(+)}) \nonumber \\*
& &-\int dt\int_{\Sigma}d^{2}x[(+)\leftrightarrow(-)].
			\label{eq:CSac}
\end{eqnarray}
As is well known this is a first class constraint system.
We have two kinds of first class constraints. First,
\begin{equation}
\Pi_{ta}^{(\pm)}\approx 0, \label{eq:constr1}
\end{equation}
where $\Pi_{ta}^{(\pm)}$ is the momentum conjugate to
$A_{t}^{(\pm)a}$.
Second,
\begin{equation}
{\cal G}_{a}^{(\pm)}\equiv\frac{L}{2}\eta_{ab}\epsilon^{ij}
(\partial_{i}A^{(\pm)b}_{j}-\partial_{j}A^{(\pm)b}_{i}+
\epsilon^{b}_{\mbox{ }cd}A^{(\pm)c}_{i}A^{(\pm)d}_{j})\approx 0,
\label{eq:constr2}
\end{equation}
which are called as Gauss law constraints.

The phase space before the reduction is parametrized by
$(A_{i}^{(\pm)a},
A_{t}^{(\pm)a},\Pi_{ta}^{(\pm)})$, whose nonvanishing
Poisson brackets
can be read off from the action (\ref{eq:CSac}):
\begin{eqnarray}
\{A_{t}^{(\pm)a}(x),\Pi_{tb}^{(\pm)}(y)\}_{P.B.}
&=&\pm\delta^{a}_{b}\delta^{2}(x,y),
\nonumber \\*
\{A_{i}^{(\pm)a}(x),A_{j}^{(\pm)b}(y)\}_{P.B.}
&=&\pm\frac{1}{L}\eta^{ab}\epsilon_{ij}\delta^{2}(x,y).
\label{eq:FTPB}
\end{eqnarray}
these are encoded in the symplectic structure of the
unreduced system:
\begin{eqnarray}
\Omega&=&\Omega^{(+)}+\Omega^{(-)}, \nonumber \\*
\Omega^{(\pm)}&=&\pm\int d^{2}x(-\frac{L}{2}\eta_{ab}\epsilon^{ij}
\delta A_{i}^{(\pm)a}\wedge\delta A_{j}^{(\pm)b}+
\delta\Pi_{ta}^{(\pm)}\wedge\delta A^{(\pm)a}_{t}),
\label{eq:FTSYM}
\end{eqnarray}
where $\delta$ denotes \lq\lq exterior derivative on the
phase space".

We will quantize the theory following the
\lq\lq reduced phase space method".
Namely, we first solve the constraints to obtain
the physical phase space,
and then we consider the quantization on the physical phase space.

The first class constraints (\ref{eq:constr1}) and (\ref{eq:constr2})
tell us that
the momentum $\Pi_{ta}^{(\pm)}$ conjugate to $A_{t}^{(\pm)a}$
vanishes and
that $A_{i}^{(\pm)a}$ be flat SO$(2,1)_{0}$ connection on $\Sigma$.
\footnote{In the gauge theory, we are often concerned with the
identity component ${\rm G}_{0}$ of the gauge group G.}
To obtain the physical phase space, we further have to take the
quotient
space modulo gauge transformations which are generated by the
first class constraints.

In our case, the generating functional of gauge transformation is
\begin{equation}
G^{(\pm)}(N,\theta)\equiv\pm\int d^{2}x
(N^{a}(x)\Pi_{ta}^{(\pm)}(x)+\theta_{a}(x){\cal G}^{(\pm)a}(x)),
\label{eq:FTGEN}
\end{equation}
where $N^{a}$ and $\theta^{a}$ in general depend on the dynamical
variables.
The infinitesimal transformation generated by (\ref{eq:FTGEN}) is
(up to terms proportional to constraints which vanishes on the
constraint surface
where the constraint equations hold):
\begin{eqnarray}
\delta_{G}A_{i}^{(\pm)a}&=&\{A_{i}^{(\pm)a},G^{(\pm)}
(N,\theta)\}_{P.B.}=
-D_{i}^{(\pm)}\theta^{a}\equiv-(\partial_{i}\theta^{a}+
\epsilon^{a}_{\mbox{ }bc}A_{i}^{(\pm)b}\theta^{c}), \nonumber \\*
\delta_{G}A_{t}^{(\pm)a}&=&\{A_{t}^{(\pm)a},G^{(\pm)}
(N,\theta)\}_{P.B.}=
N^{a}, \nonumber \\*
\delta_{G}\Pi_{ta}^{(\pm)}&=&\{\Pi_{ta}^{(\pm)},G^{(\pm)}(N,\theta)\}
_{P.B.}=0, \label{eq:GT}
\end{eqnarray}
i.e., the SO$(2,1)_{0}$ gauge transformation on $A_{i}^{(\pm)a}$
and the
shift on $A_{t}^{(\pm)a}$. In principle, $A_{t}^{(\pm)a}$
can be arbitrarily
chosen and we usually regard it as a Lagrange multiplier.
Now the resulting phase space turns out to be a direct product
${\cal M}$
of two moduli spaces ${\cal M}^{(\pm)}$ of flat SO$(2,1)_{0}$
connections
on $\Sigma$ modulo SO$(2,1)_{0}$ gauge transformations:
\begin{equation}
{\cal M}={\cal M}^{(+)}\times{\cal M}^{(-)}.
\end{equation}

Restriction of the symplectic sructure (\ref{eq:FTSYM}) to the
constraint surface
$\Pi_{ta}^{(\pm)}={\cal G}^{(\pm)a}=0$ naturally induces the
symplectic
structrue of the physical phase space ${\cal M}$.
To see this, we compute the
gauge transformation of  (\ref{eq:FTSYM}):
\begin{eqnarray}
\delta_{G}\Omega^{(\pm)}&=&
\pm\int d^{2}x[-\frac{L}{2}\epsilon^{ij}\eta_{ab}
\delta(-D_{i}^{(\pm)}\theta^{a})\wedge\delta A_{j}^{(\pm)b}\times2
+\delta\Pi^{(\pm)}_{ta}\wedge\delta N^{a}]
\nonumber \\*
&=&\int d^{2}x[-\eta_{ab}\delta\theta^{a}\wedge\delta
{\cal G}^{(\pm)b}+
\delta\Pi^{(\pm)}_{ta}\wedge\delta N^{a}].
\label{eq:GT2}
\end{eqnarray}
Because $\delta{\cal G}^{(\pm)a}=\delta\Pi_{ta}^{(\pm)}=0$ on the
constraint surface, we find $\delta_{G}\Omega=0$ and we can regard
the symplectic structure to be defined on ${\cal M}$.

Alternatively, we obtain the same result by properly fixing the gauge
and by taking the Dirac bracket.
A gauge-fixing corresponds to taking a \lq\lq
cross-section" which intersects with each
orbit of the gauge transformations once and only once,
and whose intersection
with the constraint surface is isomorphic to
the physical phase space.
Dirac bracket is given by the symplectic structure of the
\lq\lq cross section" which is
induced from the symplectic structure of the original
unconstrained system.
Taking these facts into account,
we see that the symplectic structure of
the physical phase space, i.e. eq.(\ref{eq:FTSYM})
restricted on ${\cal M}$,
should be equivalent to the symplectic structure
which is obtained by the Dirac bracket.

To parametrize ${\cal M}$, it is convenient to use holonomy of the
connection $A^{(\pm)}$ \cite{nelson}\cite{NRZ}:
\begin{equation}
h^{(\pm)}_{A}(\gamma)\equiv {\cal
P}\exp\{\int_{1}^{0}ds\dot{\gamma}^{i}(s)
A^{(\pm)}_{i}(\gamma(s))\},
\label{eq:holo}
\end{equation}
where $\gamma:[0,1]\rightarrow\Sigma$ is an arbitrary
closed curve on $\Sigma$
and the base point $x_{0}=\gamma(0)=\gamma(1)$ is assumed
to be fixed.
${\cal P}$ denotes the path ordered product,
with larger $s$ to the left.

Let us consider expressing the phase space ${\cal M}$ in terms of
(\ref{eq:holo}). Because the connection $A^{(\pm)}$ in
${\cal M}^{(\pm)}$ is flat,
the $h^{(\pm)}_{A}$ depends only on the homotopy class of the closed
curve $\gamma$. A gauge transformation of $A^{(\pm)}$
$$
A^{(\pm)}_{i}(x)\rightarrow A^{\prime(\pm)}_{i}(x)=
g^{(\pm)}(x)A^{(\pm)}_{i}(x)g^{(\pm)-1}(x)
-\partial_{i}g^{(\pm)}(x)g^{(\pm)-1}(x),\quad  g^{(\pm)}(x)\in
{\rm SO}(2,1)_{0}
$$
induces a conjugate transformation of $h^{(\pm)}_{A}$:
$$
h^{(\pm)}_{A}\rightarrow h^{(\pm)}_{A^{\prime}}=
g^{(\pm)}(x_{0})h^{(\pm)}_{A}g^{(\pm)-1}(x_{0}).
$$
Hence we can express the physical phase space as
\begin{equation}
{\cal M}^{(\pm)}=Hom(\pi_{1}(\Sigma),SO(2,1)_{0})/\sim,
\label{eq:RPS}
\end{equation}
where $Hom(A,B)$ denotes the space of group homomorphisms
$A\rightarrow B$,
$\pi_{1}(\Sigma)$ is the fundamental group of $\Sigma$,
and $\sim$ means
the equivalence under the SO$(2,1)_{0}$ conjugations.

%%%%%%%%%%%%%%%%%%%%%%%%%%%%%%%%%%%%%%

\section{Reduced Phase Space on ${\bf R}\times T^{2}$}

Now we apply the method explained in the last section to the case
where $M\approx{\bf R}\times T^{2}$.
First we look into the topological structure of the physical
phase space.

The fundamental group $\pi_{1}(T^{2})$ of a torus is generated by two
commuting generators $\alpha$ and $\beta$. The holonomies of the flat
connection $A^{(\pm)}$ therefore form a subgroup of SO$(2,1)_{0}$
generated by two commuting SO$(2,1)_{0}$ elements.
By taking an appropriate
conjugation, we know that each sub-phase spaces ${\cal M}^{(\pm)}$
consists of three subsectors ${\cal M}^{(\pm)}_{S}$,
${\cal M}^{(\pm)}_{N}$, and ${\cal M}^{(\pm)}_{T}$ \cite{ezawa}
(plus a set ${\cal M}_{0}=\{S^{(\pm)}[\alpha]=S^{(\pm)}[\beta]=0\}$
with nonzero codimensions).
\footnote{We will use the spinor representation, where the generators
of SO$(2,1)_{0}$ Lie algebra is given by pseudo-Pauli matrices
$\lambda_{a}$:
$$\lambda_{a}\lambda_{b}=\frac{1}{4}\eta_{ab}
+\frac{1}{2}\epsilon_{abc}\lambda^{c}. $$
We will henceforce denote the holonomy $h^{(\pm)}_{A}$ in the spinor
representaion by $S^{(\pm)}$.}
${\cal M}^{(\pm)}_{S}$ is parametrized by
\begin{equation}
S^{(\pm)}[\alpha]=\exp(\lambda_{2}\alpha_{\pm}),\quad
S^{(\pm)}[\beta]=\exp(\lambda_{2}\beta_{\pm}),\label{eq:ssec}
\end{equation}
with $(\alpha_{\pm},\beta_{\pm})\in ({\bf R}^{2}\backslash\{(0,0)\})
/{\bf Z}_{2}$.\footnote{${\bf Z}_{2}$ in the denominator is
generated by the internal inversion:
$(\alpha_{\pm},\beta_{\pm})\rightarrow
-(\alpha_{\pm},\beta_{\pm})$.}
Parametrization of ${\cal M}^{(\pm)}_{N}$ is\footnote{
This parametrization is different from that in \cite{ezawa}.
In fact the former includes the latter as a special case with
$\theta_{\pm}\in(-\pi/2,\pi/2)$. }
\begin{equation}
S^{(\pm)}[\alpha]=
\exp\{(\lambda_{0}\pm\lambda_{2})\cos\theta_{\pm}\},\quad
S^{(\pm)}[\beta]=\exp\{
(\lambda_{0}\pm\lambda_{2})\sin\theta_{\pm}\},\label{eq:nsec}
\end{equation}
with $\theta_{\pm}+2\pi$ being identified with $\theta_{\pm}$.
${\cal M}^{(\pm)}_{T}$ is expressed by the following
parametrization
\begin{equation}
S^{(\pm)}[\alpha]=\exp(\lambda_{0}\rho_{\pm}),\quad
S^{(\pm)}[\beta]=\exp(\lambda_{0}\sigma_{\pm}),\label{eq:tsec}
\end{equation}
where $\rho_{\pm}$ and $\sigma_{\pm}$ are periodic
with period $2\pi$.

To obtain their symplectic structures,
we have to look for  flat connections
which give the desired holonomies. Such connections are easily found.
If we use as coordinates on $T^{2}$ the periodic
coordinates $x$ and $y$
along $\alpha$ and $\beta$ with period 1, the simplest connections
are the following
\begin{eqnarray}
{\cal M}_{S}^{(\pm)}&:&A^{(\pm)}\equiv A^{(\pm)a}\lambda_{a}
=-\lambda_{2}(\alpha_{\pm}dx+\beta_{\pm}dy),
\nonumber \\*
{\cal M}_{N}^{(\pm)}&:&A^{(\pm)}=-(\lambda_{0}+\lambda_{2})
(\cos\theta_{\pm}dx+\sin\theta_{\pm}dy),
\nonumber \\*
{\cal M}_{T}^{(\pm)}&:&A^{(\pm)}=
-\lambda_{2}(\rho_{\pm}dx+\sigma_{\pm}dy).
\label{eq:SNT}
\end{eqnarray}
The symplectic structures are obtained by substituting
the above expressions
for $A^{(\pm)}$ into eq.(\ref{eq:FTSYM}).
The symplectic structure of ${\cal M}_{S}^{(\pm)}$is given by
\begin{equation}
\Omega^{(\pm)}=\mp L\delta\alpha_{\pm}\wedge\delta\beta_{\pm}.
\label{eq:ssymp}
\end{equation}
${\cal M}_{N}^{(\pm)}$ by itself does not have a
symplectic structure.
Symplectic structure of ${\cal M}_{T}^{(\pm)}$ is
\begin{equation}
\Omega^{(\pm)}=\pm L\delta\rho_{\pm}\wedge\delta\sigma_{\pm}.
\label{eq:tsymp}
\end{equation}

We would like to provide a construction in which these three
subsectors ${\cal M}^{(\pm)}_{S}$,
${\cal M}^{(\pm)}_{N}$ and ${\cal M}^{(\pm)}_{T}$ appear in one
parametrization. It turns out that this unification can
be done as in the $\Lambda=0$ case\cite{louko}.
For this purpose we first consider two commuting
SO$(2,1)_{0}$ holonomies in the following form:
\begin{eqnarray}
S^{(\pm)}[\alpha]=\exp\left[\cos\theta_{\pm}
\left\{\left(r_{\pm}+\sqrt{r_{\pm}^{\mbox{ }2}+1}\right)^{1/2}
\lambda_{0}\pm\left(-r_{\pm}+\sqrt{r_{\pm}^{\mbox{ }2}+1}\right)
^{1/2}\lambda_{2}\right\}\right], \nonumber \\*
S^{(\pm)}[\beta]=\exp\left[\sin\theta_{\pm}
\left\{\left(r_{\pm}+\sqrt{r_{\pm}^{\mbox{ }2}+1}\right)^{1/2}
\lambda_{0}\pm\left(-r_{\pm}+\sqrt{r_{\pm}^{\mbox{ }2}+1}\right)
^{1/2}\lambda_{2}\right\}\right]. \label{eq:unify}
\end{eqnarray}
The corresponding connection is given by
\begin{equation}
A^{(\pm)}=-\left\{\left(r_{\pm}+\sqrt{r_{\pm}
^{\mbox{ }2}+1}\right)^{1/2}
\lambda_{0}\pm\left(-r_{\pm}+\sqrt{r_{\pm}^{\mbox{ }2}+1}\right)
^{1/2}\lambda_{2}\right\}(\cos\theta_{\pm}dx+\sin\theta_{\pm}dy).
\label{eq:unify2}
\end{equation}
The above connection with $r_{\pm}<0$, $r_{\pm}=0$ and
$r_{\pm}>0$ give parametrization of ${\cal M}_{S}$,
${\cal M}_{N}$ and ${\cal M}_{T}$ respectively.
Relations between these new parameters $(r_{\pm},\theta_{\pm})$
and the old ones $(\alpha_{\pm},\beta_{\pm})$ for
${\cal M}^{(\pm)}_{S}$ and $(\rho_{\pm},\sigma_{\pm})$ for
${\cal M}^{(\pm)}_{T}$ are obtained by performing on
(\ref{eq:unify}) the conjugation using $\exp(
\mp\lambda_{1}\Phi_{\pm})$ with $\Phi_{\pm}=\frac{1}{2}
\ln\{|r_{\pm}|/(\sqrt{r_{\pm}^{\mbox{ }2}+1}+1)\}$:
\begin{eqnarray}
(\alpha_{\pm},\beta_{\pm})=\pm\sqrt{-2r_{\pm}}
(\cos\theta_{\pm},\sin\theta_{\pm})\quad for
\quad r_{\pm}<0, \label{eq:suni} \\
(\rho_{\pm},\sigma_{\pm})=\quad \sqrt{2r_{\pm}}
(\cos\theta_{\pm},\sin\theta_{\pm})\quad for
\quad r_{\pm}>0. \label{eq:tuni}
\end{eqnarray}
We should note that for $r_{\pm}>0$, the parameters $(r_{\pm},
\theta_{\pm})$ are subject to somewhat complicated identification
conditions due to the periodicity of $(\rho_{\pm},\sigma_{\pm})$.

Using the new parametrization, symplectic structures
(\ref{eq:ssymp}) and
(\ref{eq:tsymp}) are expressed by the unified form:
\begin{equation}
\pm L\delta r_{\pm}\wedge\delta\theta_{\pm}. \label{eq:Symp}
\end{equation}
In this expression, vanishing of the symplectic structure in
${\cal M}^{(\pm)}_{N}$ can be also explained by the fact that
$r_{\pm}$ is a constant ({\it i.e.} zero) in this subsector.

In summary, we give the topological structure of
$$
{\cal M}_{U}^{(\pm)}\equiv{\cal M}^{(\pm)}\backslash{\cal M}
^{(\pm)}_{0} ={\cal M}^{(\pm)}_{S}\cup{\cal M}^{(\pm)}_{N}\cup
{\cal M}^{(\pm)}_{T}.
$$
We should notice that the period of the parameter $\theta_{\pm}$
is $\pi$ for $r_{\pm}<0$ and $2\pi$ for $r_{\pm}\geq 0$. The
${\cal M}_{U}^{(\pm)}$ defined above therefore turns out to be
a non-Hausdorff manifold constructed by gluing together a punctured
cone ( ${\cal M}^{(\pm)}_{S}$) and a punctured torus
( ${\cal M}^{(\pm)}_{T}$) at the puncture in the one to two
fashion. The circle which serves as the glue is provided by
${\cal M}^{(\pm)}_{N}$. This structure precisely
coincides with that of the base space of cotangent bundle structure
of the phase space in the case with a vanishing
cosmological constant \cite{louko}. In the case with a
negative cosmological constant, however, the phase space
${\cal M}$ does not have a cotangent bundle structure even after
the removal of the set involving ${\cal M}^{(\pm)}_{0}$.
The phase space is represented by the direct product of two
non-Hausdorff manifolds plus
the set with nonzero codimensions.

Here we make a remark.
In obtaining the sub-phase space ${\cal M}^{(\pm)}$,
we first found out an adequate
SO$(2,1)_{0}$ holonomy and then consructed
the corresponding SO$(2,1)_{0}$
connection. In fact, this procedure involves identifying
the connections
which are related with each other by a large gauge transformation
\begin{equation}
g^{(\pm)}=\exp\{(2\pi \lambda_{0}(nx+my)\}\quad(n,m\in{\bf Z})
.\label{eq:LGT}
\end{equation}
Since SO$(2,1)_{0}$ (or SL(2,{\bf R})) is not simply connected,
this class of
gauge transformations cannot be generated by the first class
constraints (\ref{eq:FTGEN}).
Whether we should incorporate such a symmetry or not
depends on physical considerations.
If we consider the symmetry under
large gauge transformations (\ref{eq:LGT})
to be \lq\lq physically irrelevant", the result is
equivalent to that obtained when we use as a gauge group
the universal
covering group $\widetilde{\rm SL}(2,{\bf R})$ of SO$(2,1)_{0}$.
In that case the reduced phase space $\tilde{\cal M}$
is the direct product of two sub-phase spaces
$\tilde{\cal M}^{(\pm)}$:
\begin{equation}
\tilde{\cal M}^{(\pm)}=\left (\bigcup_{n,m\in{\bf Z}}
{\cal M}_{S}^{(\pm)nm}
\right)\cup\left(\bigcup_{n,m\in{\bf Z}}{\cal M}_{N}^{(\pm)nm}\right)
\cup\tilde{\cal M}_{T}^{(\pm)}
\cup\left(\bigcup_{n,m\in{\bf Z}}{\cal M}_{0}^{(\pm)nm}\right).
\end{equation}
Connection which belongs to each subsector is
\footnote{The $\tilde{\lambda}_{a}$ is the generator of
$\widetilde{\rm SL}(2,{\bf R})$ and is subject to the same
commutation relations as that of
pseudo-Pauli matrices.}
\begin{eqnarray}
{\cal M}_{S}^{(\pm)nm}&:&A^{(\pm)}=-\tilde{\lambda}_{0}2\pi(ndx+mdy)
\nonumber \\*       & &
-(\tilde{\lambda}_{2}\cos2\pi(nx+my)-\tilde{\lambda}_{1}
\sin2\pi(nx+my))
(\alpha_{\pm}dx+\beta_{\pm}dy) \nonumber \\*
{\cal M}_{N}^{(\pm)nm}&:&A^{(\pm)}=-\tilde{\lambda}_{0}2\pi(ndx+mdy)
\nonumber \\*      & &
-[\tilde{\lambda}_{0}\pm
(\tilde{\lambda}_{2}\cos2\pi(nx+my)-\tilde{\lambda}_{1}
\sin2\pi(nx+my))]
(\cos\theta_{\pm}dx+\sin\theta_{\pm}dy) \nonumber \\*
\tilde{\cal M}_{T}^{(\pm)}&:&A^{(\pm)}=-\tilde{\lambda}_{0}
(\tilde{\rho}_{\pm}dx+\tilde{\sigma}_{\pm}dy)    \nonumber \\*
{\cal M}_{0}^{(\pm)nm}&:&A^{(\pm)}=-\tilde{\lambda}_{0}2\pi(ndx+mdy),
\label{eq:EXT}
\end{eqnarray}
where the parameters $\alpha_{\pm}$, $\beta_{\pm}$ and $\theta_{\pm}$
run in the same regions as those in the SO$(2,1)_{0}$ case, but
the domain of $(\tilde{\rho}_{\pm},\tilde{\sigma}_{\pm})$ is
${\bf R}^{2}\backslash\{(2\pi n,2\pi m)| n,m\in{\bf Z}\}$.

As in the SO$(2,1)_{0}$ case we can \lq\lq unify" the sub-phase
space except
the set $\bigcup_{n,m\in{\bf Z}}{\cal M}_{0}^{(\pm)nm}$ with nonzero
codimensions. Though we cannot give coordinates which parametrize
the whole of $\tilde{\cal M}_{U}^{(\pm)}=
\tilde{\cal M}^{(\pm)}\backslash
\left(\bigcup_{n,m\in{\bf Z}}{\cal M}_{0}^{(\pm)nm}\right)$,
we can find a chart in the
neighbourhood of $A^{(\pm)}=-\tilde{\lambda}_{0}2\pi(ndx+mdy)\in
{\cal M}_{0}^{(\pm)nm}$:
\begin{equation}\begin{array}{l}
A^{(\pm)}=-\tilde{\lambda}_{0}2\pi(ndx+mdy) \\
-\left\{\left(r_{\pm}+\sqrt{r_{\pm}^{\mbox{ }2}+1}\right)^{1/2}
\tilde{\lambda}_{0} \pm\left(-r_{\pm}+
\sqrt{r_{\pm}^{\mbox{ }2}+1}\right)
^{1/2}  \right.\\
\left. \times\left(\tilde{\lambda}_{2}\cos2\pi(nx+my)-
\tilde{\lambda}_{1}\sin2\pi(nx+my)\right)\right\}
(\cos\theta_{\pm}dx+\sin\theta_{\pm}dy).\end{array}
\label{eq:unify3}
\end{equation}
The above connection with $r_{\pm}<0$, $r_{\pm}=0$ and
$r_{\pm}>0$ give parametrizations of ${\cal M}_{S}^{(\pm)nm}$,
${\cal M}_{N}^{(\pm)nm}$ and $\tilde{\cal M}_{T}^{(\pm)}$
respectively.
Relations between the old and the new parameters are
exactly the same as
those in the SO$(2,1)_{0}$ case,
provided that $(\rho_{\pm},\sigma_{\pm})$ be
replaced by $(\tilde{\rho}_{\pm}-2\pi n,
\tilde{\sigma}_{\pm}-2\pi m)$.

Since we have obtained a chart, the topological structure of
$\tilde{\cal M}_{U}^{(\pm)}$ can be read off.
Local structure of the neighbourhood of
$A^{(\pm)}=-\tilde{\lambda}_{0}2\pi(ndx+mdy)$ precisely
coinsides with
that of the neighbourhood of $A^{(\pm)}=0$ in the SO$(2,1)_{0}$ case.
Globally, $\tilde{\cal M}_{U}^{(\pm)}$ is a non-Hausdorff manifold
which is obtained by gluing infinitely many copies of a punctured
cone (${\cal M}^{(\pm)nm}_{S}$) to an infinitely many punctured plane
($\tilde{\cal M}^{(\pm)}_{T}$) at each puncture in the one-to-two
fashion. ${\cal M}^{(\pm)nm}_{N}$ serve as the glue.

%%%%%%%%%%%%%%%%%%%%%%%%%%%%%%%%%%%%

\section{Relation to Other Formalisms}

In the last section we have provided the new parametrization and
investigated the topology of the phase space of CSG.
Our choice of basic variables is, however, somewhat different from
those of the previous literatures on CSG \cite{nelson}\cite{NRZ}
\cite{carli}\cite{ezawa}. Now we will make the relations between our
variables and conventional ones transparent.

First we investigate relations to the invariants of Nelson and Regge
\cite{nelson}\cite{NRZ}. Nelson and Regge used the Wilson loop
operator in
the chiral spinor representation to parametrize the physical phase
space. Our (anti-)self-dual holonomy $S^{(\pm)}[\gamma]$ essentially
corresponds to the \lq\lq integrated connection" $S^{\pm}(\gamma)$
in \cite{NRZ}, so we can easily express the $c$-invariants of Nelson
and Regge in terms of our new variables
\begin{equation}
(c^{\pm}(\alpha),c^{\pm}(\beta))=\left\{\begin{array}{ccc}
(\cosh\frac{\alpha_{\pm}}{2},\cosh\frac{\beta_{\pm}}{2})&for&
{\cal M}_{S}^{(\pm)}\\
(\quad1\quad,\quad1\quad)&for&{\cal M}_{N}^{(\pm)}\\
(\cos\frac{\rho_{\pm}}{2},\cos\frac{\sigma_{\pm}}{2})&for&
{\cal M}_{T}^{(\pm)}.
\end{array}\right. \label{eq:NR}
\end{equation}
These can formally be rewritten in a unified fashion
\begin{equation}
(c^{\pm}(\alpha),c^{\pm}(\beta))=\left(\cos(\sqrt{\frac{r_{\pm}}{2}}
\cos\theta_{\pm}),\cos(\sqrt{\frac{r_{\pm}}{2}}\sin\theta_{\pm})
\right).
\end{equation}
Now we can give an alternative derivation of the Poisson bracket, or
the symplectic structure (\ref{eq:ssymp})(\ref{eq:tsymp}).
The Poisson bracket of $c$-invariants is given in ref.\cite{NRZ}.
After
translating into our convention, it is
\begin{equation}
\{c^{\pm}(\alpha),c^{\pm}(\beta)\}_{P.B.}=\pm\frac{1}{8L}
(c^{\pm}(\alpha\beta)-c^{\pm}(\alpha\beta^{-1})). \label{eq:NRZ}
\end{equation}
Substituting eq.(\ref{eq:NR}) into eq.(\ref{eq:NRZ}), we find, for
example for ${\cal M}_{S}^{(\pm)}$
$$
\{\cosh\frac{\alpha_{\pm}}{2},\cosh\frac{\beta_{\pm}}{2}\}_{P.B.}
=\pm\frac{1}{4L}\sinh\frac{\alpha_{\pm}}{2}\sinh\frac{\beta_{\pm}}
{2},
$$
which is equivalent to (\ref{eq:ssymp}) classically.
Similar calculation
shows the equivalence of  eq.(\ref{eq:NRZ}) to
eq.(\ref{eq:tsymp}) and
to eq.(\ref{eq:Symp}).

Next we consider the relation to the ADM formalism\cite{soda}.
In a previous paper \cite{ezawa} we have investigated relations
between the ADM formalism and CSG in detail when
$\Lambda\neq0$.
We have shown that the ADM formalism has direct correspondence
with the
$\left({\cal M}^{(+)}_{S}\times{\cal M}^{(-)}_{S}\right)$-sector
of the physical phase space of CSG. We have also shown that the ADM
variables (complex modulus $m$, conjugate momentum $p$ and
Hamiltonian $H$) can be expressed in terms of the parameters
$(\alpha,\beta,u,v)$ which were used in \cite{ezawa} to parametrize
${\cal M}^{(+)}_{S}\times{\cal M}^{(+)}_{S}$:
\begin{eqnarray}
m&=&\frac{v+i\beta\tan t}{u+i\alpha\tan t}\nonumber \\*
p&=&-iL\cot t(u-i\alpha\tan t)^{2}  \nonumber \\*
H&=&-\frac{L}{\sin t\cos t}(u\beta-v\alpha). \label{eq:ADM-CS}
\end{eqnarray}
The canonical transformation from the ADM variables to the
$(\alpha,\beta,u,v)$-variables is written as
\begin{equation}
{\rm Re}(p\delta\overline{m})-H\delta t=
2L(v\delta\alpha-u\delta\beta)+\delta F\label{eq:CAN}
\end{equation}
where
\begin{equation}
F(m_{1},m_{2},\alpha,\beta)=\frac{L\tan t}{m_{2}}|\beta-m\alpha|^{2}.
\end{equation}

So it is sufficient to show the relation between
our new paramtrization and  the old one $(\alpha,\beta,u,v)$.
By considering that these partameters are originally used to
express holonomies, it is straightforward to find
\begin{equation}
\alpha_{\pm}=\alpha\pm u \quad,\quad
\beta_{\pm}=\beta\pm v. \label{eq:O-N}
\end{equation}
Using (\ref{eq:ADM-CS}) , (\ref{eq:O-N}) and  (\ref{eq:suni}),
we find  the expressions of the ADM
variables in terms of new parameters $(r_{\pm},\theta_{\pm})
$ :
\begin{eqnarray}
m&=&\frac{e^{it}\sin\theta_{+}\sqrt{-2r_{+}}
+e^{-it}\sin\theta_{-}\sqrt{-2r_{-}}}
{e^{it}\cos\theta_{+}\sqrt{-2r_{+}}
+e^{-it}\cos\theta_{-}\sqrt{-2r_{-}} }\quad, \\
p&=&\frac{-iL}{4\sin t\cos t}\left(e^{-it}\cos\theta_{+}
\sqrt{-2r_{+}}+e^{it}\cos\theta_{-}\sqrt{-2r_{-}}\right)^{2}, \\
H&=&\frac{-L}{\sin t\cos t}\sin(\theta_{+}-\theta_{-})
\sqrt{r_{+}r_{-}}\quad,
\end{eqnarray}
which are essentially the same as those given in ref.\cite{CN}.
These new parameters  $(r_{\pm},\theta_{\pm})$ are related with
the parameters $(\alpha,
\beta,u,v)$ by an ordinary canonical transformation
\begin{eqnarray}
2L(v\delta\alpha-u\delta\beta)&=&
L(r_{+}\delta\theta_{+}-r_{-}\delta\theta_{-})-\delta V,
\nonumber \\
V(\alpha,\beta,\theta_{+},\theta_{-})&=&
2L\frac{(\alpha\sin\theta_{-}-\beta\cos\theta_{-})
(\alpha\sin\theta_{+}-\beta\cos\theta_{+})}
{\sin(\theta_{+}-\theta_{-})}\quad .\label{eq:cannon}
\end{eqnarray}
However, the canonical transformation from the ADM variables to
these new parameters is {\it singular} in the sense that it does
not contain the generating function:
\begin{equation}
{\rm Re}(p\delta\overline{m})-H\delta t=
L(r_{+}\delta\theta_{+}-r_{-}\delta\theta_{-}).
\label{eq:can2}
\end{equation}
We conjecture that this singular nature is related to the
fact that $r_{\pm}$ and therefore $p$ cannot be expressed in terms
of $(m,\overline{m},\theta_{+},\theta_{-})$ alone.

We know that the ${\cal M}_{(S,S)}$ is in 1 to 2 correspondence
with the ADM formalism \cite{ezawa}. This is originated from the
symmetry of CSG under the transformation
$$(\alpha,\beta,u,v)\rightarrow(u,v,\alpha,\beta),$$
which can be expressed in terms of the ADM formalism by
$$t\rightarrow t+\frac{\pi}{2} . $$
In the next section, we will look into this 1 to 2 correspondence
from the viewpoint of the spacetime geometry.

%%%%%%%%%%%%%%%%%%%%%%%%%%%%%%%%%%

\section{Geometrical Interpretation of the Reduced Phase Space}

In this section we try to relate a spacetime to each point in the
physical phase space. We mainly focus on the case where the
gauge group is ${\rm SO}(2,1)_{0}\times{\rm SO}(2,1)_{0}$.
We use $(x,y)$ as periodic coordinates on $T^{2}$
with period $1$. Identification conditions are therefore obvious.
Since the set involving ${\cal M}_{0}^{(\pm)}$ gives singular
universes, we only consider the
subspace ${\cal M}^{\prime}\equiv{\cal M}^{(+)}_{U}\times
{\cal M}^{(-)}_{U}$ with codimension zero,
which consists of the nine sectors. We will
denote these sectors as ${\cal M}_{(\Psi,\Phi)}\equiv
{\cal M}^{(+)}_{\Psi}\times{\cal M}^{(-)}_{\Phi}$ ($\Psi,\Phi
=S,N,T$).

As an illustration we review the spacetime construction
from ${\cal M}_{(S,S)}$ \cite{ezawa}. The simplest connection which
gives the holonomies (\ref{eq:ssec}) is given by (\ref{eq:SNT}):
\begin{equation}
A^{(\pm)}=-\lambda_{2}d\varphi_{\pm}, \quad
(\varphi_{\pm}\equiv\alpha_{\pm}x+\beta_{\pm}y).
\end{equation}
By performing a time-dependent gauge transformation
$g^{(\pm)}=e^{\mp\lambda_{0}t}$ and by extracting the triad part
\begin{equation}
E^{0}=dt,\quad E^{1}=-\sin t\frac{d\varphi_{+}+d\varphi_{-}}{2},
\quad E^{2}=-\cos t\frac{d\varphi_{+}-d\varphi_{-}}{2},
\label{eq:SSTRI}
\end{equation}
we can construct the spacetime metric
\begin{equation}
L^{-2}ds^{2}=-dt^{2}+\cos^{2}t\mbox{ }d\left(
\frac{\varphi_{+}-\varphi_{-}}{2}\right)^{2}+
\sin^{2}t\mbox{ }d\left(\frac{\varphi_{+}+\varphi_{-}}{2}
\right)^{2}.     \label{eq:SS}
\end{equation}
Parametrization of the $AdS^{3}$ which reproduces this metric is:
\begin{equation}\begin{array}{ll}
(T,X,Y,Z)=L(\sin t\cosh\frac{\varphi_{+}+\varphi_{-}}{2},&
\sin t\sinh\frac{\varphi_{+}+\varphi_{-}}{2}, \\
 &\cos t\sinh\frac{\varphi_{+}-\varphi_{-}}{2},
\cos t\cosh\frac{\varphi_{+}-\varphi_{-}}{2}). \end{array}
\label{eq:SSP}
\end{equation}

We should  remark that the periodicity condition for the
above parametrization is expressed by the identification under two
${\rm SO}(2,2)_{0}$ transformations of $(T,X,Y,Z)\in M^{2+2}$
\begin{eqnarray}
\tilde{E}[\alpha]&=& \left(\begin{array}{cccc}
                \cosh\alpha & \sinh\alpha & 0 & 0 \\
				\sinh\alpha & \cosh\alpha & 0 & 0 \\
				0 & 0 & \cosh u & \sinh u \\
				0 & 0 & \sinh u & \cosh u \\
				\end{array}  \right) \nonumber \\
\tilde{E}[\beta]&=& \left(\begin{array}{cccc}
                \cosh\beta & \sinh\beta & 0 & 0 \\
				\sinh\beta & \cosh\beta & 0 & 0 \\
				0 & 0 & \cosh v & \sinh v \\
				0 & 0 & \sinh v & \cosh v \\
				\end{array}  \right),\label{eq:so22}
\end{eqnarray}
which are given by the (anti-)self-dual SO$(2,1)_{0}$ holonomies
(\ref{eq:ssec}) through the relation
\begin{equation}
S^{(+)}\tilde{\sigma}_{\hat{a}}X^{\hat{a}}[S^{(-)}]^{-1}
=\tilde{\sigma}_{\hat{a}}\tilde{E}^{\hat{a}}_{\hat{b}}X^{\hat{b}},
\label{eq:S-V}
\end{equation}
where $\tilde{\sigma}_{\hat{a}}\equiv(2\lambda_{a},{\bf 1})$ is the
\lq\lq soldering form" in 2+2 dimensional Minkowskii space:
\begin{equation}
L^{-1}\tilde{\sigma}_{\hat{a}}X^{\hat{a}}=
L^{-1}\left(\begin{array}{cc}Y+Z & T+X \\
-T+X & -Y+Z \end{array}\right)
\in{\rm PSL}(2,{\bf Z}).
\end{equation}
The spacetime construction of Witten and Mess\cite{witte}\cite{mess},
in which we identify the spacetime $M$ with
a quotient space ${\cal F}/G$, where ${\cal F}$ is a subspace of
the anti-de Sitter space $AdS^{3}$ and $G$ is a subgroup of
SO$(2,2)$ which is specified by a point on the physical phase space,
therefore seems to be
equivalent to the standard construction explained above.

Indeed, it turns out that these two alternative constructions
give the same spacetime also to the remaining eight sectors.
We will omit the detail of its derivation and give only
parametrization in the $AdS^{3}$ which represent the
spacetime constructed from a point in each sectors.\footnote{
We always consider that
$$T^{2}-X^{2}-Y^{2}+Z^{2}=L^{2}$$
holds. The metric is obtained by substituting the
parametrization into the pseudo-Minkowski metric:
$$
ds^{2}=-dT^{2}+dX^{2}+dY^{2}-dZ^{2}.
$$}$^{,}$
\footnote{We define the following new coordinates on $T^{2}$:
$$
\eta_{\pm}\equiv x\cos\theta_{\pm}+y\sin\theta_{\pm},\quad
\zeta_{\pm}\equiv\rho_{\pm}x+\sigma_{\pm}y.
$$}\\
${\cal M}_{(N,N)}$:
\begin{equation}\begin{array}{l}
X+Z=Le^{t}, \quad (T,Y)=Le^{t}\left(
\frac{\eta_{+}-\eta_{-}}{2},\frac{\eta_{+}+\eta_{-}}{2}\right).
\end{array}\label{eq:NN}
\end{equation}
${\cal M}_{(T,T)}$:
\begin{equation}
\begin{array}{ll}
L^{-1}(T,X,Y,Z)=(\cosh t\cos\frac{\zeta_{+}-\zeta_{-}}{2},
&\sinh t\cos
\frac{\zeta_{+}+\zeta_{-}}{2},  \\ & \sinh t\sin
\frac{\zeta_{+}+\zeta_{-}}{2},
-\cosh t\sin\frac{\zeta_{+}-\zeta_{-}}{2}).
\end{array}\label{eq:TT}
\end{equation}
${\cal M}_{(T,S)}$:
\begin{equation}
\begin{array}{l}
L^{-1}(T,X,Y,Z)=   \\
\qquad \cosh t(\cosh\frac{\varphi_{-}}{2}\cos\frac{\zeta_{+}}{2},
\sinh\frac{\varphi_{-}}{2}\cos\frac{\zeta_{+}}{2},
\sinh\frac{\varphi_{-}}{2}\sin\frac{\zeta_{+}}{2},
-\cosh\frac{\varphi_{-}}{2}\sin\frac{\zeta_{+}}{2})\\
\qquad  +\sinh t(\sinh\frac{\varphi_{-}}{2}\sin\frac{\zeta_{+}}{2},
\cosh\frac{\varphi_{-}}{2}\sin\frac{\zeta_{+}}{2},
-\cosh\frac{\varphi_{-}}{2}\cos\frac{\zeta_{+}}{2},
\sinh\frac{\varphi_{-}}{2}\cos\frac{\zeta_{+}}{2}).
\end{array}\label{eq:TS}
\end{equation}
${\cal M}_{(S,N)}$:
\begin{equation}
\begin{array}{l}
L^{-1}(T,Y)=(\sin t\cosh\frac{\varphi_{+}}{2},
\cos t\sinh\frac{\varphi_{+}}{2})+\frac{\eta_{-}
(\cos t\cosh\frac{\varphi_{+}}{2}
+\sin t\sinh\frac{\varphi_{+}}{2})}{2}(-1,1),\\
{}\\
L^{-1}(Z,X)=(\cos t\cosh\frac{\varphi_{+}}{2},
\sin t\sinh\frac{\varphi_{+}}{2})-\frac{\eta_{-}
(\cos t\sinh\frac{\varphi_{+}}{2}
+\sin t\cosh\frac{\varphi_{+}}{2})}{2}(-1,1).
\end{array}\label{eq:SN}
\end{equation}
${\cal M}_{(T,N)}$:
\begin{equation}\begin{array}{l}
L^{-1}(T,Y)=(\cosh t\cos\frac{\zeta_{+}}{2},
\sinh t\sin\frac{\zeta_{+}}{2})
+\frac{\eta_{-}(\sinh t\cos\frac{\zeta_{+}}{2}
-\cosh t\sin\frac{\zeta_{+}}{2})}{2}(-1,1), \\
{} \\
L^{-1}(Z,X)=(-\cosh t\sin\frac{\zeta_{+}}{2},\sinh t\cos
\frac{\zeta_{+}}{2})
-\frac{\eta_{-}(\cosh t\cos\frac{\zeta_{+}}{2}
+\sinh t\sin\frac{\zeta_{+}}{2})}{2}(-1,1).
\end{array}\label{eq:TN}
\end{equation}
As for the other three sectors ${\cal M}_{(S,T)}$,
${\cal M}_{(N,S)}$ and ${\cal M}_{(N,T)}$, the following holds
generically. The metric obtained from a point in
${\cal M}_{(\Phi,\Psi)}$ ($\Phi\neq\Psi$) can be made into the same
form as the one obtained from ${\cal M}_{(\Psi,\Phi)}$ with the
subscripts $\pm$ replaced by $\mp$. On the other hand, the triad
and the parametrization in the former are respectively obtained by
reversing the orientation of the triad and by replacing
$Z$ with $-Z$ in the latter. This is also expected from the fact
that interchanging $S^{(+)}$ and $S^{(-)}$ in eq.(\ref{eq:S-V})
is equivalent to the conjugation of $\tilde{E}$ by
$$ P_{Z}:(T,X,Y,Z)\rightarrow(T,X,Y,-Z). $$
Taking these facts into account, we can
say that the universe obtained from a point in
${\cal M}_{(\Phi,\Psi)}$ is the \lq\lq mirror image" of that in
${\cal M}_{(\Psi,\Phi)}$.

Here we remark a few problems of the spacetime interpretation of this
type. In the above discussion we have neglected whether the action of
holonomy group is properly discontinuous. Let us consider
${\cal M}_{(T,T)}$ as an illustration. The SO$(2,2)_{0}$
holonomies in this
sector  is expressed by combining the rotations in the $(X,Y)$- and
$(T,Z)$-directions. If we consider to take the quotient of
the anti-de Sitter space, the action of the holonomy group is not
properly discontinuous. To make the action of the holonomy properly
discontinuous we have to i) take the universal covering
$\widetilde{AdS}^{3}$ of the anti-de Sitter space and ii) remove
$X=Y=0(T^{2}+Z^{2}=L^{2})$ from $\widetilde{AdS}^{3}$ and
take the universal covering of the resultant space.
After performing these prescriptions the quotient space
is made well-defined. There is, however, another problem.
To a point $(\rho_{+},\sigma_{+},\rho_{-},\sigma_{-})$
on ${\cal M}_{(T,T)}$, there correspond
infinitely many spacetimes which are obtained by
replacing $(\rho_{\pm},\sigma_{\pm})$
in the parametrization (\ref{eq:TT})
by $(\rho_{\pm}+2\pi m_{\pm},\sigma_{\pm}+2\pi n_{\pm})$.
In fact such situation is generic to the sectors
${\cal M}_{(T,\Psi)}$ and ${\cal M}_{(\Phi,T)}$.
At first sight this problem seems to be settled down by
considering the
$\widetilde{\rm SL}(2,{\bf R})\times\widetilde{\rm SL}(2,{\bf R})$
gauge theory. The problem is, however, not so simple because it is
difficult to deal with ${\cal M} _{S}^{(\pm)nm}$ or ${\cal M} _{N}
^{(\pm)nm}$. Consider
${\cal M} _{S}^{(+)n_{+}m_{+}}\times{\cal M} _{S}^{(+)n_{-}m_{-}}$
as an example. The original connection is given by (\ref{eq:EXT}).
As in the case of ${\cal M}_{(S,S)}$,
by performing the time-dependent gauge transformation
$g^{(\pm)}=e^{\mp\lambda_{0}t}$ and by extracting the triad part,
\begin{eqnarray}
E^{0}\mbox{ }&=&dt-\pi\{(n_{+}-n_{-})dx+(m_{+}-m_{-})dy\}\equiv
dt^{\prime} \nonumber \\*
\left(\begin{array}{c}E^{1} \\ E^{2}\end{array}\right)&=&
\left(\begin{array}{cc}
\cos\Theta(x,y) & -\sin\Theta(x,y) \\
\sin\Theta(x,y)  & \cos\Theta(x,y)
\end{array}\right) \times
\left(\begin{array}{c}
 -\sin t^{\prime}\frac{d\varphi_{+}+d\varphi_{-}}{2} \\
 -\cos t^{\prime}\frac{d\varphi_{+}-d\varphi_{-}}{2}
\end{array}\right),
\end{eqnarray}
where $\Theta(x,y)\equiv\pi\{(n_{+}+n_{-})x+(m_{+}+m_{-})y\}$,
we can construct the spacetime metric
\begin{equation}
ds^{2}=L^{2}\left(-dt^{\prime 2}+\sin^{2}t^{\prime}
(\frac{d\varphi_{+}+d\varphi_{-}}{2})^{2}+
\cos^{2}t^{\prime}(\frac{d\varphi_{+}-d\varphi_{-}}{2}
)^{2}\right),  \label{eq:SSEX}
\end{equation}
which seems to be the same metric as that obtained from
${\cal M}_{(S,S)}$. There is, however, an obstruction against
regarding (\ref{eq:SSEX}) and (\ref{eq:SS}) as equivalent.
In order to identify (\ref{eq:SSEX}) and (\ref{eq:SS}) we have to
regard $t^{\prime}$ in (\ref{eq:SSEX}) to be an ordinary
time function
which is single-valued on the spacetime.  As a consequence, the
gauge transformation $g^{(\pm)}=e^{\mp\lambda_{0}t}$ which we have
used to construct a nonsingular metric becomes a large gauge
transformation which relates the non-equivalent connections.
It would be more sensible to regard $t$ as a single-valued
time function
and $g^{(\pm)}=e^{\mp\lambda_{0}t}$ to be a gauge transformation
which is homotopic to the identity.
The spacetime with metric  (\ref{eq:SSEX}) is then entirely different
from the spacetime with metric  (\ref{eq:SS}) unless
$(n_{+},m_{+})=(n_{-},m_{-})$. This can be seen by being aware that
the spacetime  (\ref{eq:SSEX}) is parametrized by (\ref{eq:SSP})
with $t$ replaced by $t^{\prime}$. The spacetime (\ref{eq:SSEX}),
however, does not appear in the ordinary ADM formalism because
$(t=const.)$-hypersurface necessarily involves timelike region.

Thus it is not straightforward to deal with the
$\widetilde{\rm SL}(2,{\bf R})\times\widetilde{\rm SL}(2,{\bf R})$
gauge theory. In particular, in the case of the remaining sectors
(except $\tilde{\cal M}_{T}^{(+)}\times\tilde{\cal M}_{T}^{(-)}$)
we do not even know whether there exist any spacetimes which
correspond to a point on each sector. To elucidate the problems
on the $\widetilde{\rm SL}(2,{\bf R})\times
\widetilde{\rm SL}(2,{\bf R})$
gauge theory, more extensive analysis is longed for.

We return to the ${\rm SO}(2,2)_{0}$ gauge theory on neglecting
the problms explained above.
The eight sectors except ${\cal M}_{(S,S)}$ give spacetimes in
which each torus $T^{2}$ is timelike, so they do not correspond to
the ordinary ADM formalism.  These spacetimes are, however,
solutions of Einstein's equations as is seen from the fact that
they are constructed from the 3-dimensional anti-de Sitter space.
So we can consider
that each point in ${\cal M}^{\prime}\backslash{\cal M}_{(S,S)}$
gives such an \lq\lq exotic" spacetime \cite{louko}.
The timelike tori involved in these spacetimes
necessarily contain closed timelike curves,
which seem to be forbidden by many works\cite{hawk} to coexist
with an ordinary universe which is (at least partially) equipped with
a causal structure.  The spacetime discussed here is, however,
the  \lq\lq nether world" in which all
\lq\lq constant-time" hypersurfaces
are timelike, or the spacetime formed by gluing an ordinary universe
and the  \lq\lq nether world" using a singularity as a glue.
Such spacetimes does not seem to be supressed by \cite{hawk},
and might play an important role in the quantum gravity particularly
when we describe the epoch before and during the big bang,
as euclidean spacetimes do in the path integral approaches.
To see whether this is indeed the case, it would be necessary to
investigate the physical adequacy of these
spacetimes more rigorously.

We know that the ${\cal M}_{(S,S)}$ is in 1 to 2 correspondence
with the ADM formalism \cite{ezawa}. Now we investigate the origin
of this 1 to 2 correspondence.

We have seen that the 1 to 2 correspondence is originated from the
symmetry of ${\cal M}_{(S,S)}$ under the seemingly discrete
transformation
$$(\alpha_{+},\beta_{+},\alpha_{-},\beta_{-})\rightarrow
(\alpha_{+},\beta_{+},-\alpha_{-},-\beta_{-}). $$
This transformation is, in fact, generated by the gauge
transformation $(G^{(+)}(\pi),G^{(-)}(\pi))\equiv
(1,\exp(\pi\lambda_{0}))$ which belongs to the 1-parameter family
of transformations:
\begin{equation}
(G^{(+)}(\theta),G^{(-)}(\theta))\equiv(1,\exp(\theta\lambda_{0})).
\label{eq:1para}
\end{equation}
By performing on the connection $A^{(\pm)}=
\lambda_{2}d\varphi_{\pm}$ in ${\cal M}_{(S,S)}$ the gauge
transformation $G^{(\pm)}(\theta)$ and
a time-dependent gauge transformation
$g^{(\pm)}=e^{\mp\lambda_{0}t}$, we obtain the SO$(2,2)_{0}$
connection whose triad part is given by
\begin{eqnarray}
E^{0}_{\theta}\mbox{ }&=& dt \nonumber \\*
\left(\begin{array}{c}E^{1}_{\theta} \\
E^{2}_{\theta}\end{array}\right)&=&
\left(\begin{array}{cc}
\cos\frac{\theta}{2} & -\sin\frac{\theta}{2} \\
\sin\frac{\theta}{2} & \cos\frac{\theta}{2}\end{array}\right)
\left(\begin{array}{c}
-\sin(t+\frac{\theta}{2})\frac{d\varphi_{+}+d\varphi_{-}}{2}\\
-\cos(t+\frac{\theta}{2})\frac{d\varphi_{+}-d\varphi_{-}}{2}
\end{array}\right). \label{eq:TRI-1}
\end{eqnarray}
The transformation which lead from (\ref{eq:SSTRI}) to
(\ref{eq:TRI-1}) is the composition of a spatial rotation
and a \lq\lq time-shift"
\footnote{We can redard this \lq\lq time-shift" as a temporal
diffeomorphism, provided that a shift of the region of $t$, e.g.
from $(0,\frac{\pi}{2})$ to $(-\frac{\theta}{2},
\frac{\pi-\theta}{2})$, follows. If $t$ runs in the region
$(-\infty,\infty)$, they cannot be distinguished. In that case,
however, we have to deal with the universe with singularities on
the way of time evolution \cite{ezawa}.}
$$t\rightarrow t+\frac{\theta}{2}.$$
So we can arbitrarily choose the origin of time. If we only
consider the region which does not have singularity
on the way of time evolution, then this symmetry tells us
that we cannot distinguish the universes whose metric is given
by (\ref{eq:SS}) with the regions of time being
$(-\frac{\pi}{2},0)$ and $(0,\frac{\pi}{2})$ respectively.
We know that there are two types of singularities in the
region parametrized by (\ref{eq:SSP}), which are the lines
$t=n\pi$ and $t=(n+\frac{1}{2})\pi$. Each of the above two
universes begins with one of these singularities and ends with
the other. We can conclude that the 1 to 2 correspondence is
originated from the lack of criterion for choosing the origin of
time in our prescription to construct spacetimes.

We could explain this 1 to 2 correspondence from the viewpoint of
the SO$(2,2)_{0}$ holonomy. From one holonomy group, we can
construct two different spacetimes, e.g. the spacetimes obtained
by identifying the regions $\{T>|X|,Z>|Y|\}$
and $\{T<-|X|,Z>|Y|\}$
using the same holonomy (\ref{eq:so22}). What is peculiar to the
$\Lambda<0$ case is that we can obtain
the above spacetimes also by identifying the regions
$\{T<-|X|,Z>|Y|\}$ and $\{T>|X|,Z>|Y|\}$ using the different
(but gauge-equivalent) holonomy
\begin{eqnarray}
\tilde{E}[\alpha]&=& \left(\begin{array}{cccc}
                \cosh u & \sinh u & 0 & 0 \\
				\sinh u & \cosh u & 0 & 0 \\
				0 & 0 & \cosh\alpha & \sinh\alpha \\
				0 & 0 & \sinh\alpha & \cosh\alpha \\
				\end{array}  \right)  \nonumber \\
\tilde{E}[\beta]&=& \left(\begin{array}{cccc}
                \cosh v & \sinh v & 0 & 0 \\
				\sinh v & \cosh v & 0 & 0 \\
				0 & 0 & \cosh\beta & \sinh\beta \\
				0 & 0 & \sinh\beta & \cosh\beta \\
				\end{array}  \right).
\end{eqnarray}
We can consider this peculiar nature of the holonomy in the
anti-de Sitter case to be the origin of 1 to 2 correspondence.

%%%%%%%%%%%%%%%%%%%%%%%%%%%%%%%%%%%%

\section{Toward the Quantum Theory}

In this section we try to quantize the \lq\lq unified" phase space
${\cal M}^{\prime}$ in the SO$(2,2)_{0}$ gauge theory.
We first look into the classical
transformation property under the large diffeomorphisms and then
we construct the modular invariant quantum theory on a \lq\lq
modified phase space" ${\cal M}^{\circ}$. Finally
we investigate the relation to the quantum theory in ref
\cite{carli} which is related to the quantum ADM formalism.

As we have seen ${\cal M}^{\prime}$ does not
have a cotangent bundle structure.The most familiar
quantization where quantum states are represented by functions
of {\it coordinates} is, however, defined only when the phase space
allows a \lq\lq real polarization", whose typical example is
a cotangent bundle structure. By an artifitial
prescription we deform
the ${\cal M}^{\prime}$ into  ${\cal M}^{\circ}$ which is a cotangent
bundle on a torus.

%%%%%%%%%%%%%%%%%%%%%%%%%%%
\subsection{Modular transformations}

First we look into the behaviour of our new canonical variables
under large diffeomorphisms, in particular the inversion:
\begin{equation}
I:\mbox{ }(\alpha,\beta)\longrightarrow -(\alpha,\beta)
\qquad \left(or\quad(x,y)\rightarrow -(x,y)\right),
\end{equation}
which induces the following {\em simultaneous} transformations:
\begin{equation}
I:\mbox{ }(\theta_{\pm},r_{\pm})\longrightarrow
(\theta_{\pm}+\pi,r_{\pm}). \label{eq:inver}
\end{equation}
${\cal M}^{\prime}$ does not have a cotangent bundle
structure even after imposing this symmetry. If we perform the
following {\em artificial} prescription, however,
the resulting phase space
${\cal M}^{\circ}$ acquires a cotangent bundle structure
${\cal M}^{\circ}={\bf T}^{\ast}{\cal B}$ with the base space
${\cal B}\approx T^{2}$:\\
i) First we get rid of the $2\pi$-periodicity in $\rho_{\pm}$ and
 $\sigma_{\pm}$  which parametrize ${\cal M}_{T}^{(\pm)}$ and
make  ${\cal M}_{T}^{(\pm)}$ homeomorphic to ${\bf R}^{2}
\backslash\{(0,0)\}$.
ii) We assume that $(\theta_{+},\theta_{-}+\pi)$ can be
distinguished from $(\theta_{+},\theta_{-})$ {\em even when either
$r_{+}$ or $r_{-}$ is negative}. This involves the
assumption that the ${\cal M}_{(S,S)}$ is not in 1 to 2
correspondence but equivalent with the ADM phase space.

The \lq\lq modified" phase space ${\cal M}^{\circ}$ constructed as
above has a symplectic potential
\begin{equation}
\Theta=L(r_{+}\delta\theta_{+}-r_{-}\delta\theta_{-})\in
T^{\ast}{\cal B}
\quad(\delta\Theta=\Omega)
\end{equation}
with the base space ${\cal B}$ being parametrized by
\begin{equation}
(\theta_{+},\theta_{-})\sim(\theta_{+}+\pi,\theta_{-}+\pi)
\sim(\theta_{+}+2\pi,\theta_{-}).
\end{equation}

What is the meaning of the \lq\lq modified "
phase space ${\cal M}^{\circ}$?
The phase space of general relativity should be composed of
equivalent
classes of solutions of Einstein's equations under the
diffeomorphisms.
The double covering of ${\cal M}_{(S,S)}$ is equivalent
to the phase space of
the ADM formalism. In CSG, however, the phase space is expected to
be extended compared to the ADM phase space because CSG can contain
singularity where the spatial metric collapses.
We can regard ${\cal M}^{\circ}$
to be the phase space of the model which take such effect of CSG into
account to some extent. To obtain the {\em true} phase space of CSG,
we have to start with the $\widetilde{\rm SL}(2,{\bf R})\times
\widetilde{\rm SL}(2,{\bf R})$ gauge theory and complete
the investigation
made in the last section.

Next we investigate the behaviour under the modular group
$\Gamma={\rm PSL}(2,{\bf Z})$.
Transformations of the classical variables under the
two elementary modular transformations:
$$
S:(\alpha,\beta)\rightarrow(-\beta,\alpha),\quad
T:(\alpha,\beta)\rightarrow(\alpha+\beta,\beta),
$$
prove to be given by the following simultaneous transformations
\begin{eqnarray}
S &:&(\theta_{\pm},r_{\pm})\rightarrow
(\theta_{\pm}+\frac{\pi}{2}, r_{\pm}),\label{eq:modul} \\
T &:&(\theta_{\pm},r_{\pm})\rightarrow \left(
\frac{1}{i}\ln\left\{\frac{e^{i\theta_{\pm}}+\sin\theta_{\pm}}
{\sqrt{1+\sin2\theta_{\pm}+\sin^{2}\theta_{\pm}}}\right\},
(1+\sin2\theta_{\pm}+\sin^{2}\theta_{\pm})r_{\pm}\right).
\nonumber
\end{eqnarray}
We can show that these transformations preserve the symplectic
structure of ${\cal M}^{\prime}$ and the cotangent bundle structure
of ${\cal M}^{\circ}$.  We have only to show that the symplectic
potential $\Theta$ is also a well-defined section of $T^{\ast}
({\cal B}/ \Gamma)$, i.e. that the values of $\Theta$
before and after
the transformation coincide. As for $S$, it is straightforward.
Invariance under $T$ is demonstrated as:
\footnote{$T^{\ast}$ here does not denote a cotangent bundle
but denotes
a pull-back of a form on ${\cal M}^{\circ}$ under
the Dehn twist $T$.}
\begin{eqnarray}
T^{\ast}\Theta&=&L[T(r_{+})\delta(T(\theta_{+}))-
T(r_{-})\delta(T(\theta_{-}))]
\nonumber \\*
&=&L[(1+\sin^{2}\theta_{+}+\sin2\theta_{+})r_{+}
\delta\left\{\frac{1}{i}\ln\frac{e^{i\theta_{+}+\sin\theta_{+}}}
{\sqrt{1+\sin^{2}\theta_{+}+\sin2\theta_{+}}}\right\}
-\{(+)\leftrightarrow(-)\}]\nonumber \\*
&=&L(r_{+}\delta\theta_{+}-r_{-}\delta\theta_{-})=\Theta.
\label{eq:Dehn}
\end{eqnarray}
We therefore expect that under the assumption
made above a consistent quantum theory can be defined on the
\lq\lq fundamental region" ${\cal B}/ \Gamma$.

\subsection{Quantum Theory on the Modified Phase Space}

If we use the cotangent bundle structure ${\cal M}^{\circ}=T^{\ast}
{\cal B}$, we can construct
a representation where the quantum states are functions of $
(\theta_{+},\theta_{-})$. In the quantum theory the canonical
variables $(\theta_{\pm},r_{\pm})$ are promoted to the basic
operators which satisfy the canonical commutation relations derived
from the symplectic structure (\ref{eq:Symp}):
\begin{equation}
[\hat{\theta}_{\pm},\hat{r}_{\pm}]=\pm i\frac{1}{L}\quad, \quad
\mbox{\it zero otherwise}. \label{eq:bcom}
\end{equation}
It is probable that the action of $\hat{\theta}_{\pm}$ on the
wavefunction $\chi$ is given by multiplication
\begin{equation}
\hat{\theta}_{\pm}\chi(\theta_{+},\theta_{-})=
\theta_{\pm}\cdot\chi(\theta_{+},\theta_{-}) \quad.
\end{equation}
To determine the action of $\hat{r}_{\pm}$, however, we have to
know the integration measure or the inner product.
It would be natural that the inner product is invariant under
the modular group $\Gamma={\rm PSL}(2,{\bf Z})$. If we require
the modular invariance of  the squared modulus $\overline{\chi}\chi$
of the wave function, one of the candidates is given by
\begin{equation}
<\chi_{1}|\chi_{2}>=\int\int\frac{d\theta_{+}d\theta_{-}}
{\sin^{2}(\theta_{+}-\theta_{-})}\overline{\chi_{1}(\theta_{+},
\theta_{-})}
\chi_{2}(\theta_{+},\theta_{-}). \label{eq:MEASURE}
\end{equation}
Modular invariance of the integration measure
$\frac{d\theta_{+}d\theta_{-}}{\sin^{2}(\theta_{+}-\theta_{-})}$
can be
demonstrated by a direct calculation using (\ref{eq:modul}).

If we require the action of $\hat{r}_{\pm}$ to be self-adjoint
with respect to
the inner product (\ref{eq:MEASURE}),we find
\begin{eqnarray}
\hat{r}_{\pm}\chi(\theta_{+},\theta_{-})&=&\frac{1}{L}
\left[\mp i\frac{\partial}{\partial\theta_{\pm}}+
i\cot(\theta_{+}-\theta_{-})\right]
\chi(\theta_{+},\theta_{-}) \nonumber \\*
&=&\mp\frac{i}{L}\sin(\theta_{+}-\theta_{-})
\frac{\partial}{\partial\theta_{\pm}}\left(
\frac{1}{\sin(\theta_{+}-\theta_{-})}
\chi(\theta_{+},\theta_{-})\right).
 \label{eq:rpm}
\end{eqnarray}

To determine the modular transformation of the quantum operators,
we have to consider the issue of operator ordering seriously.
Transformation of $\hat{\theta}_{\pm}$ under $S$, $T$
and transformation
of $\hat{r}_{\pm}$ under $S$ are obtained by directly promoting the
transformation (\ref{eq:modul}) to the operator relation.
Transformation of $\hat{r}_{\pm}$ under $T$, however, involves
the operators $r_{\pm}$ and $\theta_{\pm}$
which do not commute and so
we have to determine the operator ordering.

If, for example, we require the self-adjointness of
$\hat{r}_{\pm}$ to be
preserved under the $T$-transformation, the transformation is
\begin{eqnarray}
T:\hat{r}_{\pm}&\rightarrow&
(1+\sin2\hat{\theta}_{\pm}+\sin^{2}\hat{\theta}_{\pm})^{1/2}
\hat{r}_{\pm}
(1+\sin2\hat{\theta}_{\pm}+\sin^{2}\hat{\theta}_{\pm})^{1/2}
\nonumber \\*
&=&\mp\frac{i}{L}\sin(T(\theta_{+})-T(\theta_{-}))
\frac{\partial}{\partial T(\theta_{\pm})}\left(
\frac{1}{\sin(T(\theta_{+})-T(\theta_{-}))}\right).
\end{eqnarray}
For the operators to transform in these ways, the wave function
$\chi(\theta_{+},\theta_{-})$ must be invariant,
up to a constant phase
factor, under the modular transformations.

Since the quantum theory constructed as above is defined on $
{\cal M}^{\circ}$ which is larger than the ADM phase space, we may
find a process which is not expected by quantizing the ADM
formalism.
In our quantum theory, momentum eigenstates would play an important
role because each sector is identified by the signature of
$(r_{+},r_{-})$.

\subsection{Quantum Relation between New and Old Parametrizations}

Here let us investigate the relation between two representations
in which wave functions are functions of old parameters
$(\alpha,\beta)$
and functions of new parameters $(\theta_{+},\theta_{-})$,
respectively.
We expect that such relation is given by a sort of integral
transformation.

In ref .\cite{carli}, Carlip derived the integral
transformation from quantum
ADM formalism to quantum CGG by extracting the eigenfunction of
modulus operator $\hat{m}$ in the quantum CSG and
by using it as the kernel. In our case, however, it is
difficult to perform such prescription because the relation between
old parameters $(\alpha,\beta,u,v)$ and
new parameters $(\theta_{\pm},r_{\pm}
)$ is non-polynomial as is shown by (\ref{eq:suni}).
So we use other method
which invokes the geometric quantization \cite{wood}.

We breifly explain the \lq\lq orthgonal projection"\cite{wood}
by using the situation where
a phase space ${\cal M}$ admits two transverse
real polarizations $P$ and $P^{\prime}$. The base spaces
$Q={\cal M}/P$ and
$Q^{\prime}={\cal M}/P^{\prime}$ are parametrized
by the coordinates $q^{i}$
and $q^{\prime i}$ respectively.
We denote the conjugate momenta of $q^{i}$
and $q^{\prime i}$ by $p_{i}$ and $p^{\prime}_{i}$ respectively.
Suppose that
the canonical transformation is written as
\begin{equation}
p_{i}dq^{i}= p^{\prime}_{i}dq^{\prime i}+dS(q^{i},q^{\prime i}),
\end{equation}
and that the measures of inner product in the representations based
on $Q$ and $Q^{\prime}$ are given by $\mu(q)d^{n}q$ and
$\mu^{\prime}(q^{\prime})d^{n}q^{\prime}$ respectively.
Then the integral transformation from the representation
based on $Q^{\prime}$ to that based on $Q$ is given by
\begin{equation}
\psi(q^{i})=\left(\frac{1}{2\pi}\right)^{n/2}\int_{Q^{\prime}}
\left(\frac{\mu^{\prime}}{\mu}\right)^{1/2}
d^{n}q^{\prime}\sqrt{{\rm det}\left(\frac{\partial^{2}S}
{\partial q^{j}\partial q^{\prime k}}\right)}e^{iS}
\psi(q^{\prime i}).
\label{eq:OPRO}
\end{equation}

Let us apply this formula to the double covering of ${\cal
M}_{(S,S)}$.
We replace $q^{i}$and $q^{\prime i}$ by $(\theta_{+},\theta_{-})
\in{\cal B}$ and $(\alpha,\beta)\in{\bf R}^{2}/{\bf Z}_{2}$
respectively.
The canonical transformation between
these variables is given by (\ref{eq:cannon}). If we substitute these
into (\ref{eq:OPRO}), we obtain the desired integral transformation
\begin{equation}
\chi(\theta_{+},\theta_{-})=\int d\alpha d\beta
\frac{\sqrt{-2LV}}{\pi}e^{iV}\chi(\alpha,\beta).
\end{equation}
Owing to the modular invariance of $V$, $\chi(\theta_{+},\theta_{-})$
becomes modular invariant if we require $\chi(\alpha,\beta)$ to be
modular invariant (up to a constant phase factor).
To justify this integral transformation, however, more extensive
investigation are needed as to, for example, the relations between
the operators $(\hat{\theta}_{\pm},\hat{r}_{\pm})$ and
$(\hat{\alpha},\hat{\beta},\hat{u},\hat{v})$. This is expected to
be complicated and is left to the future investigation.

Finally we shall make a digression. We could
formally apply this \lq\lq orthgonal projection"
method to the derivation of the quantum relation between the
ADM formalism
and CSG. We replace $q^{i}$and $q^{\prime i}$ by $(\alpha,\beta)$
and $(m_{1},m_{2})$ respectively. Using the canonical transformation
(\ref{eq:CAN}) we find
\begin{equation}
\chi(\alpha,\beta)=\int\frac{d^{2}m}{m_{2}^{\mbox{ }2}}
\frac{|\beta-m\alpha|}{\pi\tilde{\tau}\sqrt{m_{2}}}e^{-\frac{i}{m_{2}
\tilde{\tau}}|\beta-m\alpha|^{2}}\chi(m_{1},m_{2}),
\end{equation}
where $\tilde{\tau}\equiv\frac{1}{L}\cot t$. This expression is
different from the integral transformation
\begin{equation}
\chi(\alpha,\beta)=\int\frac{d^{2}m}{m_{2}^{\mbox{ }2}}
\frac{\beta-m\alpha}{\pi\tilde{\tau}\sqrt{m_{2}}}e^{-\frac{i}{m_{2}
\tilde{\tau}}|\beta-m\alpha|^{2}}\chi(m)
\end{equation}
which is derived by Carlip \cite{carli} by a phase factor $\exp\{i
{\rm arg}(\beta-m\alpha)\}$ in the kernel.
This is probably because we have applied the \lq\lq orthogonal
projection"
method naively to the time-dependent canonical
transformation (\ref{eq:CAN}). It would be no wonder that a
modification
is required in the case of a time-dependent canonical transformation.

%%%%%%%%%%%%%%%%%%%%%%%%%%%%%%%%%%%
\section{Discussion}

In this paper we have investigated Chern-Simons formulation
of anti-de Sitter gravity on ${\bf R}\times T^{2}$ with an emphasis
on the properties of the whole phase space. In particular, we have
shown that the nine sectors which appeared in ref.\cite{ezawa} are in
fact not disconnected but are mutually connected to form
the \lq\lq unified" phase space ${\cal M}^{\prime}$, which is
a direct product of two copies of a non-Hausdorff manifold, plus a
set
with nonzero codimensions. We have also seen that each point
on ${\cal M}^{\prime}$ corresponds to a spacetime
(or spacetimes) which is a solution of  Einstein's equations
with a negative cosmological constant.
In order to quantize this theory in a conventional fashion,
we have made an artifitial prescription to modify ${\cal
M}^{\prime}$.
${\cal M}^{\circ}$ obtained in this way enjoys a cotangent
bundle structure which is preserved under the modular
transformations.
This property is convenient to the one who want to construct
a  modular invariant quantum theory. Though somewhat formally,
we have also given the relation between our new quantum theory and
the quantum theory which was given in \cite{ezawa}
and which is closely related to the spinor representation
\cite{carli}
of the ADM formalism.

While we have investigated CSG on ${\bf R}\times T^{2}$
considerably extensively, there remain many issues to be resolved
in order to complete the analysis. We will list some of these issues.

In giving the spacetime interpretation to  ${\cal M}^{\prime}$,
which is obtained by regarding the gauge group as
${\rm SO}(2,1)_{0}\times{\rm SO}(2,1)_{0}$,
we have seen that infinitely many spacetimes correspond to
each point on the $(T,\Phi)$- or the $(\Phi,T)$-sectors.
 Thus we expect $\tilde{\cal M}_{U}^{(+)}\times
\tilde{\cal M}_{U}^{(-)}$, which is obtained by choosing
$\widetilde{\rm SL}(2,{\bf R})\times\widetilde{\rm SL}(2,{\bf R})$
as the gauge group, to be more suitable to the spacetime
interpretation. Relating spacetimes to the all points on
$\tilde{\cal M}_{U}^{(+)}\times\tilde{\cal M}_{U}^{(-)}$,
however, requires a considerable exertion. Moreover,
the choice as to whether we identify the different points on
$\tilde{\cal M}_{U}^{(+)}\times\tilde{\cal M}_{U}^{(-)}$ which
give the same spacetime or not changes the structure of
the \lq\lq physical phase space" drastically.\footnote{
We can see a similar example in ref.\cite{ezawa2}, which deals with
the de Sitter case.}
After we construct the \lq\lq true" phase  space in CSG,
we have to quantize this phase space using the geometric
quantization scheme. Though the quantum theory which we have
given in \S 6 is constructed on the modified phase space, it
is based on the method of the geometric quantization and so
we can probably extend the prescription developed in \S 6 to
the complete quantization of the \lq\lq true" phase space.
To accomplish this task it is necessary to find out the
complete quantum relation between the old and the new
parametrizations.

We should note that the spacetimes we have given are not the
unique ones constructed from the points in ${\cal M}^{\prime}$.
It is because the gauge group SO$(2,2)_{0}$ (or $
\widetilde{\rm SL}(2,{\bf R})\times\widetilde{\rm SL}(2,{\bf R})$)
is in fact larger than the semi-direct product of the 2+1
dimensional local Lorentz group and the group of diffeomorphisms
\cite{louko}\cite{unruh}.
For illustration, we consider ${\cal M}_{(S,S)}$. By choosing
time dependent gauge transformations other than that giving the
spacetime (\ref{eq:SS}), we can construct various spacetimes.
There are for example Louko-Marolf-type universe \cite{louko}
and Unruh-Newbury-type universe \cite{unruh}
in which timelike tori appear.
Though these spacetimes coincide with one another in the region
where the ADM is
well-defined ($T>|X|$, $Z>|Y|$), their behaviors in the other
region vary considerably by the choice of gauge.
At present there seems to be no
criterion for choosing the most relevant gauge.

In \S 5 we have investigated the origin of the 1 to 2 correspondence
with the ADM formalism. In the de Sitter case, there exists
1 to $\infty$ correspondence\cite{ezawa2},
whose origin also have to be elucidated.
We consider that this 1 to 2 correspondence is

closely related to the fact
that the SO(3,1) gauge group is in fact larger than the
semi-direct product of the (2+1)-local Lorentz group and
the diffeomorphism group, in particular when the triad is
degenerate.

To extract instructions on the (3+1)-dimensional quantum
gravity, it is necessary to compare the reduced phase space
method which has been discussed in this paper to Dirac's
quantization method\cite{dirac}.
Witten has applied this Dirac's quantization
in the de Sitter case with the help of geometric
quantization \cite{witte3}. It is worth investigating
whether Witten's prescription can be extended to
the anti-de Sitter case.

\vskip2.5cm

\noindent Acknowledgments

I would like to thank Prof. K. Kikkawa, Prof. H. Itoyama and H.
Kunitomo for helpful discussions and
careful readings of the manuscript.

%%%%%%%%%%%%%%%%%%%%%%%%%%%%%%%%%%%%%%%%%%%%%%%%%%%%%%%%%%%%%%%%%%%

\end{document}